\documentclass[aps,pra,groupedaddress,twocolumn,citeautoscript]{revtex4-1}

\usepackage{graphicx}
\usepackage{stmaryrd}
\usepackage{color}
\usepackage{hyperref}
\usepackage{amsmath}

\definecolor{blue_n}{rgb}{0.,0.3,0.5}

\hypersetup{
backref=true, 
pagebackref=true,
hyperindex=true, 
colorlinks=true, 
breaklinks=true, 
citecolor = blue_n,
urlcolor= blue_n,
linkcolor= black, 
pdftitle={Single-photon emitters at telecom wavelengths in silicon}, 
pdfauthor={
W. Redjem, A. Durand, T. Herzig, A. Benali, 
S. Pezzagna, J. Meijer, A. Yu. Kuznetsov, 
H. S. Nguyen, S. Cueff,
I. Robert-Philip, B. Gil, D. Caliste, P. Pochet, 
M. Abbarchi, V. Jacques, A. Dr\'eau, G. Cassabois } 
}

\begin{document}

\title{Single artificial atoms in silicon emitting at telecom wavelengths} 

\author{W. Redjem$^{1}$}
\thanks{Contributed equally to this work.}
\author{A. Durand$^{1}$}
\thanks{Contributed equally to this work.}
\author{T. Herzig$^{2}$}
\author{A. Benali$^{3}$}
\author{S. Pezzagna$^{2}$}
\author{J. Meijer$^{2}$}
\author{A. Yu. Kuznetsov$^{4}$}
\author{H. S. Nguyen$^{5}$}
\author{S. Cueff$^{5}$}
\author{J.-M. G\'erard$^6$}
\author{I. Robert-Philip$^{1}$}
\author{B. Gil$^{1}$}
\author{D.~Caliste$^{6}$}
\author{P. Pochet$^{6}$}
\author{M. Abbarchi$^{3}$}
\author{V. Jacques$^{1}$}
\author{A. Dr\'eau$^{1}$}
\email[]{anais.dreau@umontpellier.fr}
\author{G. Cassabois$^{1}$}

\affiliation{$^1$Laboratoire Charles Coulomb, Universit\'e de Montpellier and CNRS, 34095 Montpellier, France}
\affiliation{$^2$Division of Applied Quantum Systems, Felix-Bloch Institute for Solid-State Physics, University Leipzig, Linn\'estra{\ss}e 5, 04103 Leipzig, Germany}
\affiliation{$^3$CNRS, Aix-Marseille Universit\'e, Centrale Marseille, IM2NP, UMR 7334, Campus de St. J\'er\^ome, 13397 Marseille, France}
\affiliation{$^4$Department of Physics, University of Oslo, NO-0316 Oslo, Norway}
\affiliation{$^5$Institut des Nanotechnologies de Lyon-INL, UMR CNRS 5270, CNRS, Ecole Centrale de Lyon, Ecully, France}
\affiliation{$^6$Department of Physics, IRIG, Univ. Grenoble Alpes and CEA, F-38000 Grenoble, France.}

\maketitle

Given its unrivaled potential of integration and scalability, silicon is likely to become a key platform for large-scale quantum technologies. Individual electron-encoded artificial atoms either formed by impurities~\cite{he_two-qubit_2019} or quantum dots~\cite{watson_programmable_2018,maurand_cmos_2016} have emerged as a promising solution for silicon-based integrated quantum circuits. However, single qubits featuring an optical interface needed for large-distance exchange of information~\cite{hensen_loophole-free_2015} have not yet been isolated in such a prevailing semiconductor. Here we show the isolation of single optically-active point defects in a commercial silicon-on-insulator wafer implanted with carbon atoms. These artificial atoms exhibit a bright, linearly polarized single-photon emission at telecom wavelengths suitable for long-distance propagation in optical fibers. Our results demonstrate that despite its small bandgap ($\simeq 1.1$ eV) \textit{a priori} unfavorable towards such observation~\cite{weber_quantum_2010}, silicon can accommodate point defects optically isolable at single scale, like in wide-bandgap semiconductors~\cite{aharonovich_solid-state_2016}. 
This work opens numerous perspectives for silicon-based quantum technologies, from integrated quantum photonics to quantum communications~\cite{wehner_quantum_2018} and metrology.

Capitalizing on the great success of the microelectronics industry, silicon is undoubtedly a promising platform for deploying large-scale quantum technologies. Silicon-based electrical qubits associated either to individual dopants~\cite{he_two-qubit_2019} or to gate-defined quantum dots~\cite{watson_programmable_2018,maurand_cmos_2016}, have already been used to demonstrate the elementary building blocks towards scalable integrated quantum circuits.
Besides requiring operation in a dilution fridge, those matter qubits are still not able to remotely exchange quantum information at long distances because they are not interfaced with optical light.~On the other side, photonic qubits at telecom wavelengths can be generated inside silicon by probabilistic non-linear optical processes~\cite{qiang_large-scale_2018}.
Even if they are adapted to long-distance propagation, those photons are not coupled to matter quantum systems, thus limiting the implementation of scalable silicon quantum photonics~\cite{silverstone_silicon_2016}. 
Another type of quantum systems that can fill the gap but is still lacking in this industry-friendly platform are optically-active spin defects, that combines an optical interface with a solid-state medium to encode quantum information~\cite{awschalom_quantum_2018}.
A \textit{sine qua non} condition to develop such spin-photon interfaces is first to demonstrate that individual point defects can be optically isolated in silicon.

\begin{figure}[t]
\begin{centering}
\includegraphics[width=1.\columnwidth]{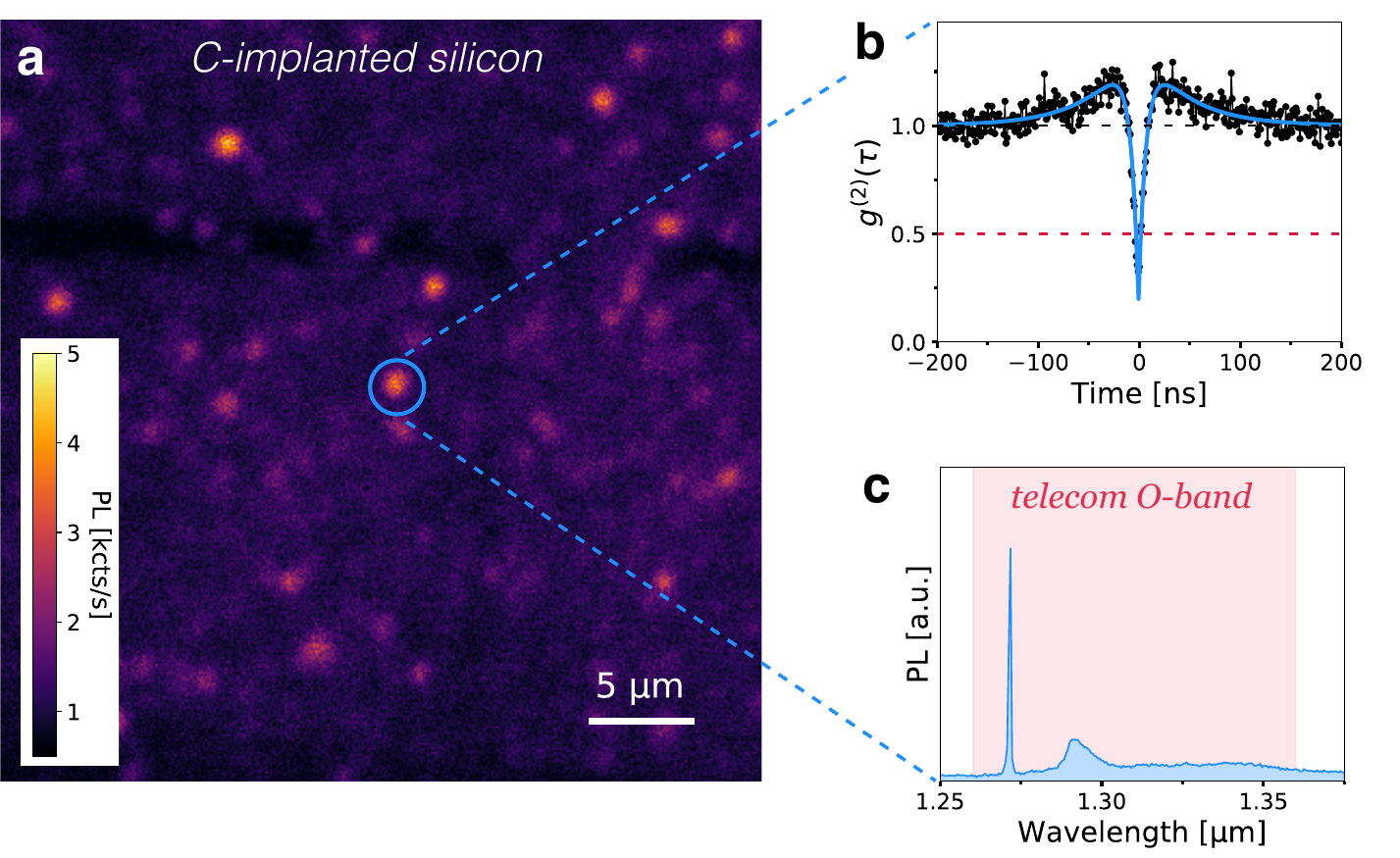}
\caption{\textbf{Single optically-active artificial atoms in silicon.} 
\textbf{a}, PL raster scan of the carbon-implanted SOI wafer recorded at 10 K for a green laser excitation power $P=10$~$\mu$W. 
\textbf{b}, Second-order correlation function $g^{(2)}(\tau)$ recorded on the isolated PL spot marked with the circle in \textbf{a}. 
The unicity of the emitter is evidenced by the clear antibunching effect at zero delay: ${g^{(2)}(0)<0.5}$. 
There is no background correction. 
%\textit{The signal-to-background ratio is $\sim 4$. }
The solid line is data fitting with a three-level model describing the dynamics of optical cycles (see Methods). 
\textbf{c}, PL spectrum recorded for the same defect, showing emission in the telecom O-band (shaded red area).}
\label{fig:Fig1}
\end{centering}
\end{figure}

Deep-level point defects in semiconductors feature electronic states buried inside the bandgap of the material~\cite{aharonovich_solid-state_2016}.~Under laser illumination, some of these defects behave as artificial solid-state atoms offering multiple quantum functionalities~\cite{awschalom_quantum_2018}. 
Diamond has been the pioneering platform for the detection and control of individual point defects~\cite{gruber_scanning_1997}, which can be used as single-photon sources for quantum cryptography~\cite{beveratos_single_2002}, spin-photon interfaces for quantum networks~\cite{hensen_loophole-free_2015} or high-sensitivity quantum sensors~\cite{gross_real-space_2017} to name a few.
In the last years, optically-active point defects have been detected at single scale in a large variety of wide-bandgap materials~\cite{aharonovich_solid-state_2016},
thus offering to extend such experiments on novel platforms better suited to industrial processes.
Despite its ubiquity in the microelectronics industry, silicon was until now still absent of the list. Owing to its narrow bandgap ($\simeq 1.1$ eV), it was an open question whether or not this semiconductor could accommodate optically-active single defects with electronic states well-isolated from its valence and conduction bands~\cite{weber_quantum_2010}. 
Here we give a positive answer to this question by demonstrating the isolation of individual photoluminescent point defects hosted in the silicon matrix.
These defects exhibit several striking features including (i) linearly polarized single-photon emission at telecom wavelengths, (ii) long-term photostability, and (iii) high brightness.

The studied sample consists of a commercial 220-nm-thick silicon-on-insulator (SOI) wafer implanted with $36$-keV carbon ions at a fluence of $5\times10^{13}$~cm$^{-2}$ in view of creating carbon-related defects~\cite{davies_optical_1989,beaufils_optical_2018} (see Methods for details). The photoluminescence (PL) response of this carbon-implanted SOI sample was studied with a scanning confocal microscope optimized for near-infrared spectroscopy at cryogenic temperature. Optical excitation was performed far above the silicon bandgap with a 532-nm continuous laser focused onto the sample through a high numerical aperture microscope objective (NA = 0.85). The PL signal was collected by the same objective and directed either to a spectrometer or to fiber-coupled infrared single-photon detectors featuring a quantum efficiency $\eta_{\mathrm{det}} = 10\%$  (see Methods for details). 

\begin{figure}[t]
\begin{centering}
\includegraphics[width=1.\columnwidth]{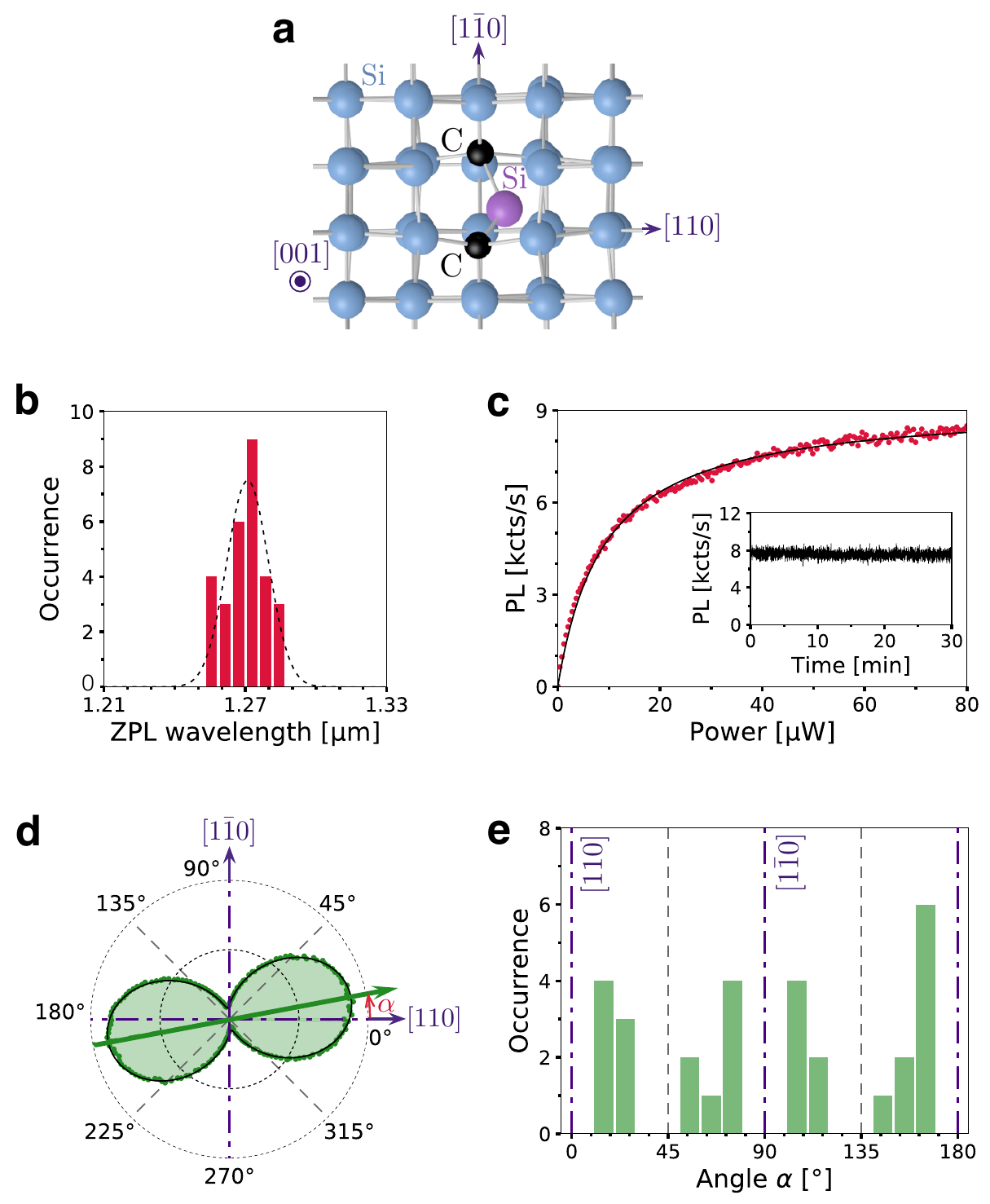}
\caption{\textbf{Photophysics of single G-centers in silicon. }
\textbf{a}, Atomic structure of the G-center consisting of an interstitial silicon atom (purple) bridging two adjacent carbon atoms (black). Atomic positions are obtained through density functional theory calculations (see Methods for details).
\textbf{b}, Histogram of the ZPL wavelengths for a set of 29 individual G-centers.
\textbf{c}, PL saturation curve. The solid line is data fitting with a simple saturation function, leading to a saturation power $P_{sat} \simeq 9$~$\mu$W. Inset: PL time-trace recorded while saturating the optical transition. The bin time is 57~ms per point. 
\textbf{d}, Polarization diagram of the PL emission indicating an emission dipole with a projection (green arrow) on the (001)-oriented SOI sample pointing at $\alpha \simeq 10^{\circ}$ from the $[110]$ crystal axis. 
\textbf{e}, Histogram of the angle $\alpha$ measured for 29 individual G-centers. 
Data shown in \textbf{c-d} are taken for the same G-center as for Fig. 1b-c. All experiments are performed at 10 K.}
\label{fig:Fig2}
\end{centering}
\end{figure}

Figure~1a depicts a typical PL raster scan of the silicon sample recorded at 10~K, which reveals bright diffraction-limited PL spots. 
To demonstrate that they correspond to individual artificial atoms, the resulting photon emission statistics was systematically analyzed by measuring the second-order correlation function $g^{(2)}(\tau)$ with two infrared single photon detectors mounted in a Hanbury-Brown and Twiss configuration. As shown in Figure 1b, a clear antibunching is observed at zero delay, ${g^{(2)}(0)\approx0.3}$, which is a fingerprint of single-photon emission from an individual quantum emitter. The deviation from an ideal single-photon emission, {\it i.e.} $g^{(2)}(0) = 0$, is well explained by residual background photons and detector dark counts~\cite{beveratos_room_2002}. 
This constitutes the first observation of single optically-active point defects embedded in the silicon lattice.

\begin{figure*}[t]
\begin{center}
\includegraphics[width=16cm]{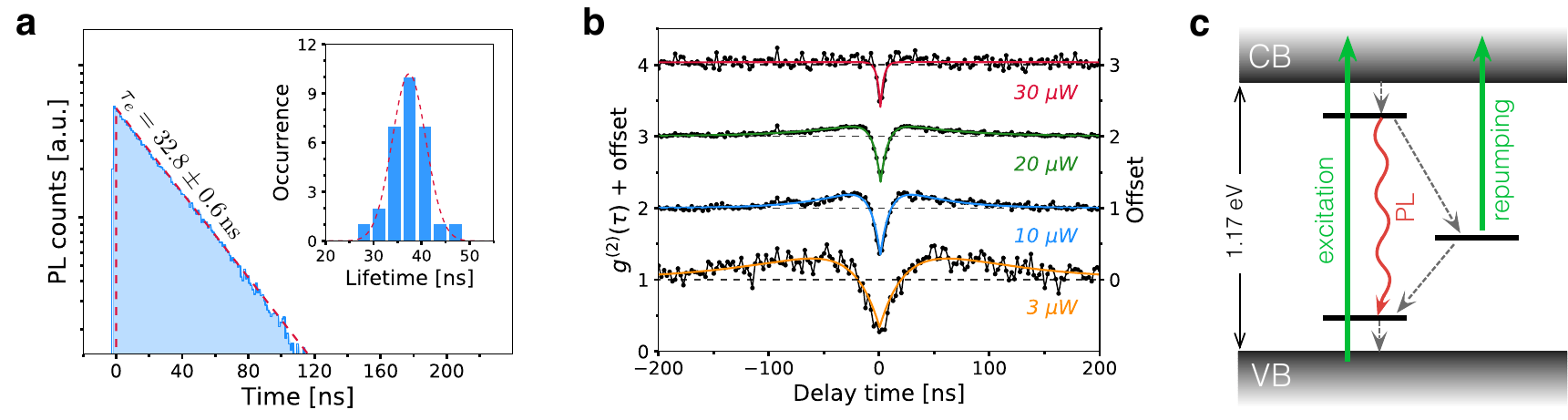}
\caption{\textbf{Dynamics of optical cycles.}
\textbf{a}, Time-resolved PL decay (in semi-log scale) recorded under optical excitation with 50-ps laser pulses at $532$~nm. 
The excited-level lifetime $\tau_e$ is extracted through data fitting with a single exponential function. Inset: Histogram of lifetime measurements for 29 individual G-centers.  
\textbf{b}, Measurements of the second-order correlation function $g^{(2)}(\tau)$ with increasing laser excitation power (from bottom to top). 
Solid-lines are data fitting with a three-level model.
\textbf{c}, The vanishing bunching effect at high laser powers is explained by introducing a photo-induced repumping mechanism from the metastable level to the excited level via the conduction band (see Methods for details). 
Data shown in \textbf{a-b} are taken for the same G-center as for Fig. 2c-d.}
\label{fig:Fig3}
\end{center}
\end{figure*}

Besides their isolation at single scale, a further asset of such emitters is that their PL spectrum falls into the O-band of optical telecommunications, with a sharp zero-phonon-line (ZPL) at a wavelength around $1.27$~$\mu$m followed by a broad phonon sideband (Fig.~1c). The Debye-Waller factor, corresponding to the proportion of photons emitted into the ZPL, reaches about $30\%$. Such a PL spectrum is very similar to the one produced by dense ensembles of G-centers in silicon~\cite{davies_optical_1989,beaufils_optical_2018,Thewalt}. 
We thus attribute the observed single emitters in the telecom O-band to individual G-centers in silicon. Even though the precise structure of this defect still remains under debate~\cite{docaj_three_2012, wang_g-centers_2014,timerkaeva_structural_2018}, it is now widely accepted as being made of a pair of carbon atoms bridging an interstitial silicon atom~\cite{song_bistable_1988, capaz_theory_1998} (Fig. 2a). 
The C-Si-C complex has a conformation resembling a water molecule, with a (C-Si-C) angle of $126 ^{\circ}$ (see Methods for details). 
We note that PL spectra recorded for a set of 29 single G-centers indicate variations of the ZPL wavelength (Fig. 2b).  
A Gaussian fit of the corresponding distribution gives a standard deviation of $9 \pm 1$ nm.
This observation likely results from local strain-induced effects, as already reported for large ensembles of G-centers under uniaxial stress~\cite{tkachev_piezospectroscopic_1978}.

The G-center shows several appealing properties in the prospect of developing integrated single-photon sources in silicon. Besides an emission at telecom wavelengths, individual G-centers exhibit a long-term stability under optical illumination, showing neither blinking nor bleaching effects (Fig. 2c Inset), and are robust against repeated thermal cycles from 10~K to room temperature. Furthermore, this defect is particularly bright under green laser illumination. A typical PL saturation curve recorded for a single G-center is shown in Figure 2c. A maximum counting rate around $8000$~cts/s can be obtained at saturation, despite (i) the high refractive index of silicon ($n\simeq 3.5$) leading to total internal reflection at the silicon-air interface and (ii) the low quantum efficiency of our infrared single-photon detectors ($\eta_{\mathrm{det}} = 10\%$). Significant improvements of the detected PL signal could be easily obtained in future by using  superconducting single photon detectors and/or by integrating G-centers into silicon photonic structures~\cite{schmitt_observation_2015,cueff_tailoring_2019}. Last, the G-center features a linearly-polarized single-photon emission. Figure~2d shows a typical emission polarization diagram recorded by rotating a polarizer in front of the single-photon detectors. The observed PL modulation with a contrast close to unity indicates that the G-center behaves as a single emitting dipole, whose projection on the (001)-oriented silicon surface is here lying at $\alpha \simeq 10 ^{\circ}$ from the [110] crystal axis. A systematic analysis over a set of 29 single G-centers reveals that the emission dipoles are pointing in specific directions distributed across four subgroups, which exclude the [110] and [1$\bar{1}$0] axis (Fig.~2e). These preferential directions are attributed to the different orientations of the G-center in the silicon matrix~\cite{lee_optical_1982}.

To gain insights into the dynamics of optical cycles, time-resolved PL measurements were performed with a pulsed-laser excitation at 532~nm. As shown in Figure~3a, the PL decay is well fitted by a mono-exponential function providing the lifetime $\tau_e$ of the excited level. Measurements performed for several single G-centers indicate fluctuations of the excited-level lifetime (Fig. 3a Inset). The resulting statistics is well-described by a Gaussian distribution with a mean value $\tau_e= 35.8 \pm 0.2$~ns and a standard deviation $\sigma_{\tau_e}=3.5 \pm 0.2$~ns. No correlation could be found between the lifetime of the excited level and neither the shift of the ZPL wavelength nor the emission dipole orientation. We note a significant discrepancy with the value recently reported for ensembles of G-centers ($\tau_e \sim 6$~ns)~\cite{beaufils_optical_2018}. This variation could originate from changes in the dynamics involving different charge states or defect conformations~\cite{song_bistable_1988} when G-centers are either isolated or inside dense ensembles. A similar effect was recently reported for nitrogen-vacancy (NV) defects in diamond, whose excited-level lifetime decreases from $\sim 12$~ns for a single defect hosted in an ultrapure diamond matrix down to $\sim 4$~ns for dense NV ensembles in a diamond sample heavily doped with nitrogen~\cite{Manson_2018}. 

Further informations can be obtained by analyzing the evolution of the second-order correlation function $g^{(2)}(\tau)$ with the laser excitation power (Fig.~3b). A photon bunching effect ($g^{(2)}(\tau) > 1$) is observed at low power, revealing that optical cycles involve non-radiative relaxation processes through a metastable level~\cite{beveratos_room_2002}. Interestingly, this effect gradually disappears while increasing the laser power, until the $g^{(2)}(\tau)$ function mimics the dynamics of a pure two-level system (top trace in Fig. 3b). 
This peculiar behavior can be explained by considering a photo-induced repumping mechanism from the metastable level to the excited level (Fig. 3c). This process likely contributes to the high brightness of single G-centers under green laser illumination.

The isolation of single point defects emitting at telecom wavelengths in silicon provides numerous opportunities for integrated quantum technologies.
First, these single-photon emitters could be integrated in photonic circuits fabricated by industrial-grade and CMOS-compatible processes~\cite{silverstone_silicon_2016}.~Full processing of photonic qubits from their deterministic generation to their measurement with integrated single-photon detectors could then be achieved on a silicon chip.~Coupling these individual emitters to highly mature silicon-based nanomechanical systems~\cite{riedinger_remote_2018} could also open new pathways in hybrid quantum optomechanics.~Finally, a remaining challenge is to assess if the internal spin degree of freedom of a single G-center in silicon can be monitored  through optical detection of the magnetic resonance, as reported a long time ago for ensembles of G-centers~\cite{lee_optical_1982}. 
The resulting spin-photon interface at telecom wavelengths would open large prospects for long-distance fibered-based quantum networks~\cite{wehner_quantum_2018} and quantum metrology.\\

\bibliography{article_Redjem_Durand_biblio.bib}

\begin{thebibliography}{10}
\expandafter\ifx\csname url\endcsname\relax
  \def\url#1{\texttt{#1}}\fi
\expandafter\ifx\csname urlprefix\endcsname\relax\def\urlprefix{URL }\fi
\providecommand{\bibinfo}[2]{#2}
\providecommand{\eprint}[2][]{\url{#2}}

\bibitem{he_two-qubit_2019}
\bibinfo{author}{He, Y.} \emph{et~al.}
\newblock \bibinfo{title}{A two-qubit gate between phosphorus donor electrons
  in silicon}.
\newblock \emph{\bibinfo{journal}{Nature}} \textbf{\bibinfo{volume}{571}},
  \bibinfo{pages}{371} (\bibinfo{year}{2019}).

\bibitem{watson_programmable_2018}
\bibinfo{author}{Watson, T.~F.} \emph{et~al.}
\newblock \bibinfo{title}{A programmable two-qubit quantum processor in
  silicon}.
\newblock \emph{\bibinfo{journal}{Nature}} \textbf{\bibinfo{volume}{555}},
  \bibinfo{pages}{633--637} (\bibinfo{year}{2018}).

\bibitem{maurand_cmos_2016}
\bibinfo{author}{Maurand, R.} \emph{et~al.}
\newblock \bibinfo{title}{A {CMOS} silicon spin qubit}.
\newblock \emph{\bibinfo{journal}{Nature Communications}}
  \textbf{\bibinfo{volume}{7}}, \bibinfo{pages}{13575} (\bibinfo{year}{2016}).

\bibitem{hensen_loophole-free_2015}
\bibinfo{author}{Hensen, B.} \emph{et~al.}
\newblock \bibinfo{title}{Loophole-free {Bell} inequality violation using
  electron spins separated by 1.3 kilometres}.
\newblock \emph{\bibinfo{journal}{Nature}} \textbf{\bibinfo{volume}{526}},
  \bibinfo{pages}{682--686} (\bibinfo{year}{2015}).

\bibitem{weber_quantum_2010}
\bibinfo{author}{Weber, J.~R.} \emph{et~al.}
\newblock \bibinfo{title}{Quantum computing with defects}.
\newblock \emph{\bibinfo{journal}{Proceedings of the National Academy of
  Sciences}} \textbf{\bibinfo{volume}{107}}, \bibinfo{pages}{8513--8518}
  (\bibinfo{year}{2010}).

\bibitem{aharonovich_solid-state_2016}
\bibinfo{author}{Aharonovich, I.}, \bibinfo{author}{Englund, D.} \&
  \bibinfo{author}{Toth, M.}
\newblock \bibinfo{title}{Solid-state single-photon emitters}.
\newblock \emph{\bibinfo{journal}{Nature Photonics}}
  \textbf{\bibinfo{volume}{10}}, \bibinfo{pages}{631--641}
  (\bibinfo{year}{2016}).

\bibitem{wehner_quantum_2018}
\bibinfo{author}{Wehner, S.}, \bibinfo{author}{Elkouss, D.} \&
  \bibinfo{author}{Hanson, R.}
\newblock \bibinfo{title}{Quantum internet: {A} vision for the road ahead}.
\newblock \emph{\bibinfo{journal}{Science}} \textbf{\bibinfo{volume}{362}},
  \bibinfo{pages}{eaam9288} (\bibinfo{year}{2018}).

\bibitem{qiang_large-scale_2018}
\bibinfo{author}{Qiang, X.} \emph{et~al.}
\newblock \bibinfo{title}{Large-scale silicon quantum photonics implementing
  arbitrary two-qubit processing}.
\newblock \emph{\bibinfo{journal}{Nature Photonics}}
  \textbf{\bibinfo{volume}{12}}, \bibinfo{pages}{534} (\bibinfo{year}{2018}).

\bibitem{silverstone_silicon_2016}
\bibinfo{author}{Silverstone, J.~W.}, \bibinfo{author}{Bonneau, D.},
  \bibinfo{author}{O'Brien, J.~L.} \& \bibinfo{author}{Thompson, M.~G.}
\newblock \bibinfo{title}{Silicon {Quantum} {Photonics}}.
\newblock \emph{\bibinfo{journal}{IEEE Journal of Selected Topics in Quantum
  Electronics}} \textbf{\bibinfo{volume}{22}}, \bibinfo{pages}{390--402}
  (\bibinfo{year}{2016}).

\bibitem{awschalom_quantum_2018}
\bibinfo{author}{Awschalom, D.~D.}, \bibinfo{author}{Hanson, R.},
  \bibinfo{author}{Wrachtrup, J.} \& \bibinfo{author}{Zhou, B.~B.}
\newblock \bibinfo{title}{Quantum technologies with optically interfaced
  solid-state spins}.
\newblock \emph{\bibinfo{journal}{Nature Photonics}}
  \textbf{\bibinfo{volume}{12}}, \bibinfo{pages}{516--527}
  (\bibinfo{year}{2018}).

\bibitem{gruber_scanning_1997}
\bibinfo{author}{Gruber, A.} \emph{et~al.}
\newblock \bibinfo{title}{Scanning confocal optical microscopy and magnetic
  resonance on single defect centers}.
\newblock \emph{\bibinfo{journal}{Science}} \textbf{\bibinfo{volume}{276}},
  \bibinfo{pages}{2012--2014} (\bibinfo{year}{1997}).

\bibitem{beveratos_single_2002}
\bibinfo{author}{Beveratos, A.} \emph{et~al.}
\newblock \bibinfo{title}{Single {Photon} {Quantum} {Cryptography}}.
\newblock \emph{\bibinfo{journal}{Physical Review Letters}}
  \textbf{\bibinfo{volume}{89}}, \bibinfo{pages}{187901}
  (\bibinfo{year}{2002}).

\bibitem{gross_real-space_2017}
\bibinfo{author}{Gross, I.} \emph{et~al.}
\newblock \bibinfo{title}{Real-space imaging of non-collinear antiferromagnetic
  order with a single-spin magnetometer}.
\newblock \emph{\bibinfo{journal}{Nature}} \textbf{\bibinfo{volume}{549}},
  \bibinfo{pages}{252--256} (\bibinfo{year}{2017}).

\bibitem{davies_optical_1989}
\bibinfo{author}{Davies, G.}
\newblock \bibinfo{title}{The optical properties of luminescence centres in
  silicon}.
\newblock \emph{\bibinfo{journal}{Physics Reports}}
  \textbf{\bibinfo{volume}{176}}, \bibinfo{pages}{83--188}
  (\bibinfo{year}{1989}).

\bibitem{beaufils_optical_2018}
\bibinfo{author}{Beaufils, C.} \emph{et~al.}
\newblock \bibinfo{title}{Optical properties of an ensemble of {G}-centers in
  silicon}.
\newblock \emph{\bibinfo{journal}{Physical Review B}}
  \textbf{\bibinfo{volume}{97}}, \bibinfo{pages}{035303}
  (\bibinfo{year}{2018}).

\bibitem{beveratos_room_2002}
\bibinfo{author}{Beveratos, A.} \emph{et~al.}
\newblock \bibinfo{title}{Room temperature stable single-photon source}.
\newblock \emph{\bibinfo{journal}{The European Physical Journal D}}
  \textbf{\bibinfo{volume}{18}}, \bibinfo{pages}{191--196}
  (\bibinfo{year}{2002}).

\bibitem{Thewalt}
\bibinfo{author}{Chartrand, C.} \emph{et~al.}
\newblock \bibinfo{title}{Highly enriched $^{28}\mathrm{Si}$ reveals remarkable
  optical linewidths and fine structure for well-known damage centers}.
\newblock \emph{\bibinfo{journal}{Phys. Rev. B}} \textbf{\bibinfo{volume}{98}},
  \bibinfo{pages}{195201} (\bibinfo{year}{2018}).

\bibitem{docaj_three_2012}
\bibinfo{author}{Docaj, A.} \& \bibinfo{author}{Estreicher, S.~K.}
\newblock \bibinfo{title}{Three carbon pairs in {Si}}.
\newblock \emph{\bibinfo{journal}{Physica B: Condensed Matter}}
  \textbf{\bibinfo{volume}{407}}, \bibinfo{pages}{2981--2984}
  (\bibinfo{year}{2012}).

\bibitem{wang_g-centers_2014}
\bibinfo{author}{Wang, H.}, \bibinfo{author}{Chroneos, A.},
  \bibinfo{author}{Londos, C.~A.}, \bibinfo{author}{Sgourou, E.~N.} \&
  \bibinfo{author}{Schwingenschlegl, U.}
\newblock \bibinfo{title}{G-centers in irradiated silicon revisited: {A}
  screened hybrid density functional theory approach}.
\newblock \emph{\bibinfo{journal}{Journal of Applied Physics}}
  \textbf{\bibinfo{volume}{115}}, \bibinfo{pages}{183509}
  (\bibinfo{year}{2014}).

\bibitem{timerkaeva_structural_2018}
\bibinfo{author}{Timerkaeva, D.}, \bibinfo{author}{Attaccalite, C.},
  \bibinfo{author}{Brenet, G.}, \bibinfo{author}{Caliste, D.} \&
  \bibinfo{author}{Pochet, P.}
\newblock \bibinfo{title}{Structural, electronic, and optical properties of the
  {C}-{C} complex in bulk silicon from first principles}.
\newblock \emph{\bibinfo{journal}{Journal of Applied Physics}}
  \textbf{\bibinfo{volume}{123}}, \bibinfo{pages}{161421}
  (\bibinfo{year}{2018}).

\bibitem{song_bistable_1988}
\bibinfo{author}{Song, L.~W.}, \bibinfo{author}{Zhan, X.~D.},
  \bibinfo{author}{Benson, B.~W.} \& \bibinfo{author}{Watkins, G.~D.}
\newblock \bibinfo{title}{Bistable defect in silicon:the
  interstitial-carbon-substitutional-carbon pair}.
\newblock \emph{\bibinfo{journal}{Physical Review Letters}}
  \textbf{\bibinfo{volume}{60}}, \bibinfo{pages}{460--463}
  (\bibinfo{year}{1988}).

\bibitem{capaz_theory_1998}
\bibinfo{author}{Capaz, R.~B.}, \bibinfo{author}{Dal~Pino, A.} \&
  \bibinfo{author}{Joannopoulos, J.~D.}
\newblock \bibinfo{title}{Theory of carbon-carbon pairs in silicon}.
\newblock \emph{\bibinfo{journal}{Physical Review B}}
  \textbf{\bibinfo{volume}{58}}, \bibinfo{pages}{9845--9850}
  (\bibinfo{year}{1998}).

\bibitem{tkachev_piezospectroscopic_1978}
\bibinfo{author}{Tkachev, V.~D.} \& \bibinfo{author}{Mudryi, A.~V.}
\newblock \bibinfo{title}{Piezospectroscopic effect on zero-phonon luminescence
  lines of silicon}.
\newblock \emph{\bibinfo{journal}{Journal of Applied Spectroscopy}}
  \textbf{\bibinfo{volume}{29}}, \bibinfo{pages}{1485--1491}
  (\bibinfo{year}{1978}).

\bibitem{schmitt_observation_2015}
\bibinfo{author}{Schmitt, S.~W.}, \bibinfo{author}{Sarau, G.} \&
  \bibinfo{author}{Christiansen, S.}
\newblock \bibinfo{title}{Observation of strongly enhanced photoluminescence
  from inverted cone-shaped silicon nanostructures}.
\newblock \emph{\bibinfo{journal}{Scientific Reports}}
  \textbf{\bibinfo{volume}{5}}, \bibinfo{pages}{17089} (\bibinfo{year}{2015}).

\bibitem{cueff_tailoring_2019}
\bibinfo{author}{Cueff, S.} \emph{et~al.}
\newblock \bibinfo{title}{Tailoring the {Local} {Density} of {Optical} {States}
  and {Directionality} of {Light} {Emission} by {Symmetry} {Breaking}}.
\newblock \emph{\bibinfo{journal}{IEEE Journal of Selected Topics in Quantum
  Electronics}} \textbf{\bibinfo{volume}{25}}, \bibinfo{pages}{1--7}
  (\bibinfo{year}{2019}).

\bibitem{lee_optical_1982}
\bibinfo{author}{Lee, K.~M.}, \bibinfo{author}{O'Donnell, K.~P.},
  \bibinfo{author}{Weber, J.}, \bibinfo{author}{Cavenett, B.~C.} \&
  \bibinfo{author}{Watkins, G.~D.}
\newblock \bibinfo{title}{Optical {Detection} of {Magnetic} {Resonance} for a
  {Deep}-{Level} {Defect} in {Silicon}}.
\newblock \emph{\bibinfo{journal}{Physical Review Letters}}
  \textbf{\bibinfo{volume}{48}}, \bibinfo{pages}{37--40}
  (\bibinfo{year}{1982}).

\bibitem{Manson_2018}
\bibinfo{author}{Manson, N.~B.} \emph{et~al.}
\newblock \bibinfo{title}{$\mathrm{NV}^-$-$\mathrm{N}^+$ pair centre in 1b
  diamond}.
\newblock \emph{\bibinfo{journal}{New Journal of Physics}}
  \textbf{\bibinfo{volume}{20}}, \bibinfo{pages}{113037}
  (\bibinfo{year}{2018}).

\bibitem{riedinger_remote_2018}
\bibinfo{author}{Riedinger, R.} \emph{et~al.}
\newblock \bibinfo{title}{Remote quantum entanglement between two
  micromechanical oscillators}.
\newblock \emph{\bibinfo{journal}{Nature}} \textbf{\bibinfo{volume}{556}},
  \bibinfo{pages}{473--477} (\bibinfo{year}{2018}).

\end{thebibliography}

\end{document}